\documentclass[runningheads]{llncs}
\usepackage{graphicx}

\usepackage{cite}
\usepackage{amsmath,amssymb,amsfonts}
\usepackage{algorithmic}
\usepackage{graphicx}
\usepackage{textcomp}
\usepackage{xcolor}
\usepackage{listings}
\usepackage{url}
\usepackage{color}
\usepackage[flushleft]{threeparttable}
\usepackage[inline]{enumitem}
\usepackage{multirow}
\usepackage[font=footnotesize]{subfig}
\usepackage{diagbox}
\usepackage{booktabs}
\definecolor{lightgray}{rgb}{0.95, 0.95, 0.95}
\definecolor{darkgray}{rgb}{0.4, 0.4, 0.4}
\definecolor{editorGray}{rgb}{0.95, 0.95, 0.95}
\definecolor{editorOcher}{rgb}{0.5, 0.26, 0} % #FF7F00 -> rgb(239, 169, 0)
\definecolor{editorGreen}{rgb}{0, 0.5, 0} % #007C00 -> rgb(0, 124, 0)
\definecolor{orange}{rgb}{1,0.45,0.13}		
\definecolor{olive}{rgb}{0.17,0.59,0.20}
\definecolor{brown}{rgb}{0.69,0.31,0.31}
\definecolor{purple}{rgb}{0.38,0.18,0.81}
\definecolor{lightblue}{rgb}{0.1,0.57,0.7}
\definecolor{lightred}{rgb}{1,0.4,0.5}
\usepackage{upquote}
\usepackage{listings}
\lstdefinelanguage{CSS}{
  keywords={color,background-image:,margin,padding,font,weight,display,position,top,left,right,bottom,list,style,border,size,white,space,min,width, transition:, transform:, transition-property, transition-duration, transition-timing-function},	
  sensitive=true,
  morecomment=[l]{//},
  morecomment=[s]{/*}{*/},
  morestring=[b]',
  morestring=[b]",
  alsoletter={:},
  alsodigit={-}
}

\lstdefinelanguage{JavaScript}{
  morekeywords={typeof, new, true, false, catch, function, return, null, catch, switch, var, if, in, while, do, else, case, break},
  morecomment=[s]{/*}{*/},
  morecomment=[l]//,
  morestring=[b]",
  morestring=[b]'
}

\lstdefinelanguage{HTML5}{
  language=html,
  sensitive=true,	
  alsoletter={<>=-},	
  morecomment=[s]{<!-}{-->},
  tag=[s],
  otherkeywords={
  >,
	<!DOCTYPE,
  </html, <html, <head, <title, </title, <style, </style, <link, </head, <meta, />,
	</body, <body,
	</div, <div, </div>, 
	</p, <p, </p>,
	</script, <script,
  <canvas, /canvas>, <svg, <rect, <animateTransform, </rect>, </svg>, <video, <source, <iframe, </iframe, </video>, <image, </image>, <header, </header, <article, </article
  },
  ndkeywords={
  =,
  charset=, src=, id=, width=, height=, style=, type=, rel=, href=,
  fill=, attributeName=, begin=, dur=, from=, to=, poster=, controls=, x=, y=, repeatCount=, xlink:href=,
  margin:, padding:, background-image:, border:, top:, left:, position:, width:, height:, margin-top:, margin-bottom:, font-size:, line-height:,
  transform:, -moz-transform:, -webkit-transform:,
  animation:, -webkit-animation:,
  transition:,  transition-duration:, transition-property:, transition-timing-function:,
  }
}

\lstdefinestyle{htmlcssjs} {%
  basicstyle={\footnotesize\ttfamily\small},   
  frame=b,
  xleftmargin={0.75cm},
  numbers=left,
  stepnumber=1,
  firstnumber=1,
  numberfirstline=true,	
  identifierstyle=\color{black},
  keywordstyle=\color{blue}\bfseries,
  ndkeywordstyle=\color{editorGreen}\bfseries,
  stringstyle=\color{editorOcher}\ttfamily,
  commentstyle=\color{editorGreen}\ttfamily,
  language=HTML5,
  alsolanguage=JavaScript,
  alsodigit={.:;},	
  tabsize=2,
  showtabs=false,
  showspaces=false,
  showstringspaces=false,
  extendedchars=true,
  breaklines=true,
  literate=%
  {Ö}{{\"O}}1
  {Ä}{{\"A}}1
  {Ü}{{\"U}}1
  {ß}{{\ss}}1
  {ü}{{\"u}}1
  {ä}{{\"a}}1
  {ö}{{\"o}}1
}
\lstdefinestyle{py} {%
language=python,
literate=%
*{0}{{{\color{lightred}0}}}1
{1}{{{\color{lightred}1}}}1
{2}{{{\color{lightred}2}}}1
{3}{{{\color{lightred}3}}}1
{4}{{{\color{lightred}4}}}1
{5}{{{\color{lightred}5}}}1
{6}{{{\color{lightred}6}}}1
{7}{{{\color{lightred}7}}}1
{8}{{{\color{lightred}8}}}1
{9}{{{\color{lightred}9}}}1,
basicstyle=\footnotesize\ttfamily, % Standardschrift
numbers=left,               % Ort der Zeilennummern
numbersep=5pt,              % Abstand der Nummern zum Text
tabsize=4,                  % Groesse von Tabs
extendedchars=true,         %
breaklines=true,            % Zeilen werden Umgebrochen
keywordstyle=\color{blue}\bfseries,
frame=b,
commentstyle=\color{editorGreen}\itshape,
stringstyle=\color{editorOcher}\ttfamily, % Farbe der String
showspaces=false,           % Leerzeichen anzeigen ?
showtabs=false,             % Tabs anzeigen ?
xleftmargin=17pt,
framexleftmargin=17pt,
framexrightmargin=5pt,
framexbottommargin=4pt,
showstringspaces=false,      % Leerzeichen in Strings anzeigen ?
}%

\lstdefinestyle{stack} {%
language=python,
basicstyle=\footnotesize\ttfamily, % Standardschrift
numbers=left,               % Ort der Zeilennummern
numbersep=10pt,              % Abstand der Nummern zum Text
tabsize=4,                  % Groesse von Tabs
extendedchars=true,         %
breaklines=true,            % Zeilen werden Umgebrochen
frame=b,
showspaces=false,           % Leerzeichen anzeigen ?
showtabs=false,             % Tabs anzeigen ?
xleftmargin=17pt,
framexleftmargin=17pt,
framexrightmargin=5pt,
framexbottommargin=4pt,
showstringspaces=false,      % Leerzeichen in Strings anzeigen ?
}%
\begin{document}
\graphicspath{{img/}}
\title{PoEW:Encryption as Consensus and Enabling Data Compression Services?}

\author{}
\institute{}
\author{Chong Guan}
% \inst{1}\orcidID{0000-1111-2222-3333} \and
%Second Author\inst{2,3}\orcidID{1111-2222-3333-4444} \and
%Third Author\inst{3}\orcidID{2222--3333-4444-5555}}
%\institute{Princeton University, Princeton NJ 08544, USA \and
%Springer Heidelberg, Tiergartenstr. 17, 69121 Heidelberg, Germany
\institute{
\email{771427477@qq.com}}
\maketitle              % typeset the header of the contribution
%

%!TEX root = ../main.tex
\begin{abstract}

Proof-of-Work (PoW) is a fundamental method in decentralized digital networks for establishing consensus on a shared ledger. By requiring network participants to solve a mathematical puzzle, PoW maintains network integrity. However, PoW has raised environmental concerns due to its significant energy consumption.

This paper introduces Proof-of-Encryption-Work (PoEW), a novel PoW consensus mechanism that repurposes computational power to address the challenge of encryption-based data compression.  PoEW uses an exhaustive key search as the PoW puzzle.  Given a lengthy plaintext and a fixed ciphertext, the corresponding key is derived.  Since the plaintext is much longer than both the key and the ciphertext, this process compresses the plaintext to the key.  This data compression is computationally intensive, while decompression is straightforward.

PS: I do not plan to submit this paper at the moment, as I recently lost my job.
Perhaps a few months later, after I find a new position, I will submit it.
This paper mainly presents a preliminary idea and still requires further development and technical details.
If you find the idea interesting, feel free to extend it. Discussions and feedback are also welcome.

\end{abstract}

%\begin{abstract}
%The abstract should briefly summarize the contents of the paper in
%15--250 words.

%\keywords{First keyword  \and Second keyword \and Another keyword.}
%\end{abstract}
%
%
%
\section{Introduction}%
\label{sec:introduction}

Proof-of-Work (PoW)~\cite{nakamoto2008bitcoin} is a fundamental method used in decentralized digital networks to achieve agreement on a shared record of information, known as a ledger. This process requires network participants to perform a computationally intensive task to solve a complex mathematical problem. The participant who successfully solves this problem is granted the authority to add a new batch of verified transactions, called a block, to the existing chain of blocks, thus updating the ledger. This not only confirms transactions but also makes it very difficult for any single party to tamper with past records, as it would require redoing the computational work for all subsequent blocks.

The complexity of the computational task acts as a security measure against malicious actions, such as attempting to spend the same digital currency multiple times. By making the validation and addition of new blocks resource-intensive, PoW ensures that maintaining the network's integrity is economically more advantageous than trying to compromise it. This security framework has been crucial for the establishment and operation of many cryptocurrencies and blockchain platforms.

Despite its success in securing networks, the PoW consensus mechanism has notable limitations. The substantial energy consumption associated with the intensive computations has led to environmental concerns and questions about its long-term viability~\cite{sapra2023impact}. Additionally, as more participants join the network, the difficulty of the computational problems tends to increase, resulting in even greater energy demands. This has motivated the exploration of alternative consensus mechanisms that aim to offer similar levels of security and decentralization with enhanced efficiency.

In response to these challenges, this paper presents a novel Proof-of-Work consensus mechanism, PoEW(Proof of Encryption Work), which offers the distinct advantage of being adaptable for data compression. By examining the design and capabilities of this new approach, this work aims to contribute to the ongoing development of consensus mechanisms within blockchain technology, particularly in its potential to integrate data compression functionalities.

In this paper, we adopt encryption as the complex mathematical problem in PoW instead of the SHA-256.The mathematical problem here is to find a key to make every block of the ciphertext begins with a number of zero bits. And when the zero bits is long enough, we can save the key and non-zero part of the ciphertext as the compressed data.

\section{Encryption as Consensus}%
\label{sec:encryptAsConsensus}

Consensus is a crucial component of blockchain technology, as it ensures that all participants in the network agree on the current state of the ledger without relying on a central authority. One of the most well-known consensus mechanisms is Proof of Work (PoW), which relies heavily on hash operations. In PoW, miners compete to solve complex mathematical puzzles that require significant computational power, typically involving repeatedly hashing input data until a specific condition is met. This process is both time-consuming and resource-intensive, making it difficult for any single entity to dominate the network. Once a valid solution is found, it is broadcast to the network for verification, and the first miner to solve the puzzle earns the right to add a new block to the blockchain and receive a reward. 

\begin{figure}[!t]
  \begin{center}
    \includegraphics[width=0.95\textwidth]{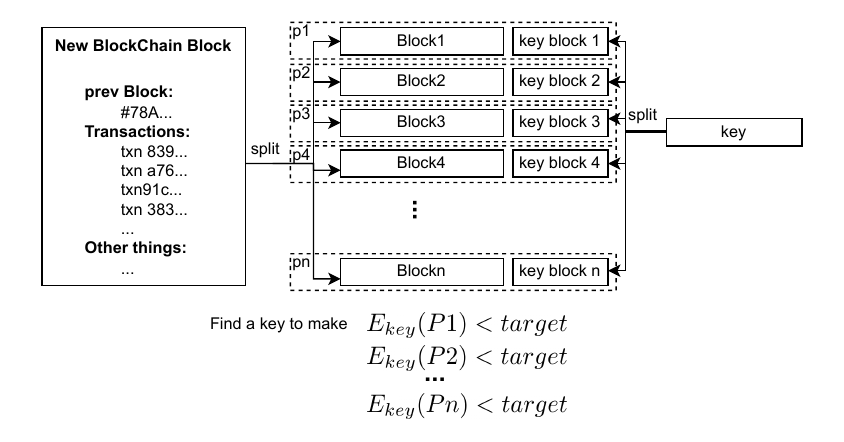}
  \end{center}
  \caption{The puzzle of PoEW}\label{fig:PoEW}
\end{figure}

In this paper, we introduce a Proof-of-Work (PoW) variant, termed PoEW, which employs the concept of an exhaustive key search from block cipher cryptography as its computational puzzle. The fundamental idea of PoEW is illustrated in Figure~\ref{fig:PoEW}.  A new blockchain block and a cryptographic key are each divided into n segments.  Corresponding segments from the block and the key are combined directly to create a series of plaintext blocks. The PoEW puzzle involves finding a key that, when used to encrypt each of these plaintext blocks, produces ciphertext where the first several bits of each encrypted block are zero.

To demonstrate PoEW, we use the Data Encryption Standard (DES) for encryption.  DES uses a 64-bit key (with 56 bits being the effective key strength) and operates on 64-bit plaintext blocks.  A Bitcoin block header, for comparison, is 80 bytes (640 bits).  In our PoEW construction, the difficulty target is also 64 bits, aligning with the ciphertext block size, thereby eliminating the need for a nonce.  Bitcoin's original PoW combines a 32-bit difficulty target and a 32-bit nonce, resulting in 64 bits.  In PoEW, since we no longer need the nonce, the length of the block header does not change.  Combining the block header and the key yields a 704-bit input, which can be divided into 11 blocks, as shown in Figure~\ref{fig:PoEW}.  Consequently, miners must find a key that satisfies the condition  $E_{key}(Pn) < target$ to solve the PoW puzzle.  Although important, a detailed discussion of the mechanism for adjusting the difficulty target is beyond the current scope of this paper.  Future work will explore this mechanism in depth.

\section{How to Compress Data With Encryption}%

Data compression reduces the size of data, allowing for its original form to be recovered. While encryption also transforms data, it typically maintains the same length. However, encryption involves a separate component—the key—which has a fixed length.

\begin{figure}[!t]
  \begin{center}
    \includegraphics[width=0.85\textwidth]{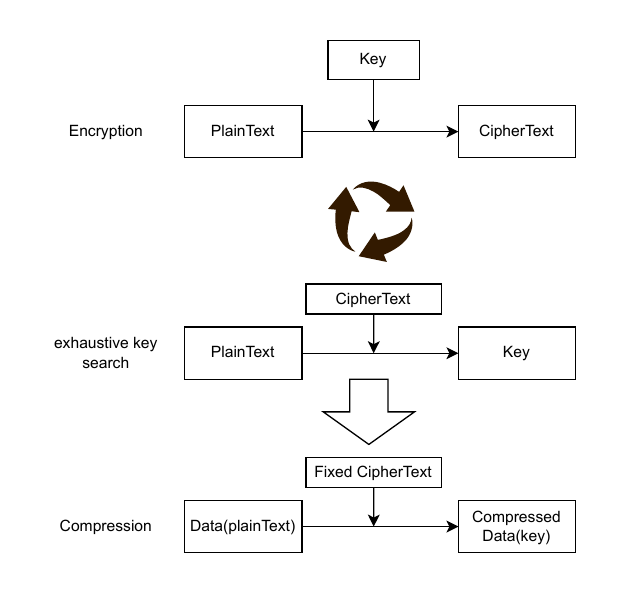}
  \end{center}
  \caption{From Encryption to Data Compression}\label{fig:eCompress}
\end{figure}

As illustrated in the first part of Figure~\ref{fig:eCompress}, encryption uses the key to select a specific transformation of the plaintext into ciphertext. By reinterpreting this process as an exhaustive key search, we can view it as a transformation that converts a long input (the plaintext) into a shorter output (the key). We propose that this perspective can be adapted for data compression.

Consider the Caesar Cipher, a simple encryption method. If we designate "AAA" as the fixed ciphertext, then "BBB" is compressed to "B", "CCC" to "C", and so on. In this instance, the compression effectively reduces three repeated characters to one. However, this method is limited to compressing data with repeated characters.  For broader applicability, the challenge lies in identifying a ciphertext whose corresponding plaintext can represent all possible data types. For example, compressing an audio file would ideally require a ciphertext such that every data block of the song could be encrypted to that ciphertext with a specific key.  Finding such a ciphertext is a complex problem that we leave for future research.

A more practical approach involves fixing only a portion of the ciphertext. For instance, we could require the ciphertext to be 8 characters long and begin with "AAA." The compressed data would then consist of the key and the unfixed portion of the ciphertext. While the compression ratio is less optimal, this method still achieves data reduction.

\section{Compress Data With Blockchain} % (fold)
\label{sec:Compression in Consensusj}

% section Compression in Consensusj (end)

In a blockchain context, PoEW does not compress data when the difficulty target is low. If we treat a 640-bit block header as the original data, and use DES encryption, and the difficulty target requires the first n bits to be zero, the compressed data would be 
$(64-n)11 + 56 = 760 - 11n$ bits. 

Data compression occurs when $ 760 - 11n < 640$ , or when $ n > 120/11$ . This represents a lower bound, as the block header is not the primary data we aim to compress. In general, achieving such a difficulty target is not feasible. However, in blockchain systems like Bitcoin, the aggregate hash rate is substantial.  The Bitcoin network's current hash rate is approximately $ 8.45 x 10^23$ hashes per second~\cite{website:blockchain}.  If we were to repurpose this computational power for a brute-force DES key search (and optimistically equate one DES key trial to one SHA-256 hash), the worst-case time to crack DES would be:

% \begin{equation}

\begin{center}
$
    T = \frac{2^{56}}{\text{Hashrate}} = \frac{7.2 \times 10^{16}}{8.452 \times 10^{23}} \approx 8.52 \times 10^{-8} \ \text{seconds}
    $
\end{center}
    
% \end{equation}

Take the DES example, we need to solve 11 such key exhaustive searching. 
It is possible that for one of the 11 plainText blocks, we can not find any key to meet the difficulty target requirement. To avoid such a cases, we need how much computational power is still unknown and waiting for future research to solve it.

\section{Related Work}%
\label{sec:related_work}

Several approaches have been proposed to improve the Proof-of-Work (PoW) consensus mechanism, addressing concerns such as energy consumption, fairness, and system scalability. Lasla et al.~\cite{lasla2022green} introduced Green-PoW, an energy-efficient variant of PoW that selectively limits mining competition in successive rounds, achieving nearly a 50\% reduction in energy usage. Similarly, Szalachowski et al.~\cite{szalachowski2019strongchain} proposed StrongChain, a collaborative PoW mechanism that enhances security and fairness by incentivizing miners to aggregate their efforts transparently, reducing centralization risks. In an effort to balance performance and security, Interleave~\cite{sun2019interleaving} presents a hybrid consensus model that interleaves PoW with Proof-of-Stake (PoS), demonstrating improved throughput and resistance to attacks. To address reward distribution fairness, a low-variance PoW system was proposed in~\cite{bazzanella2023bitcoin}, requiring miners to find multiple nonces, thereby mitigating the volatility of mining rewards. Furthermore, Skyscraper~\cite{bouvier2025skyscraper} explores fast hashing using large prime numbers, presenting an efficient PoW construction that significantly increases hash throughput while preserving cryptographic strength, offering potential gains in both performance and energy efficiency.

\section{Discussion and Conclusion}%
\label{sec:discussion}

In this paper, we explore a new approach to Proof-of-Work (PoW) consensus.  PoW is a mechanism where solving a computational puzzle is required to validate transactions and add new blocks to a blockchain.

We build on the idea that encryption can serve as this computational puzzle, with the added benefit of facilitating data compression.  Current data compression methods resemble a form of encryption where the key can be easily derived from the original data and the compressed output.  Consequently, a robust, unbreakable encryption algorithm has the potential to achieve superior data compression.

However, using secure encryption for data compression introduces a significant challenge: the compressed data cannot be easily derived.  Instead, it requires an exhaustive search of all possibilities, which is impractical for typical data compression applications.  Fortunately, blockchain systems inherently involve a substantial amount of computational effort.

This paper introduces Proof-of-Encryption-Work (PoEW), a novel PoW consensus mechanism that repurposes this computational power to address the challenge of encryption-based data compression. 

% \input{tex/7-Conclusion.tex}
\bibliographystyle{splncs04}
%\bibliography{mybibliography}
\bibliography{main}
\end{document}